\documentclass[amstex,showpacs,floats,floatfix,superscriptaddress,aps,pra,twocolumn]{revtex4}
\usepackage{amssymb}
\usepackage{amsmath}
\usepackage{calc}
\usepackage{graphicx}
\usepackage{bm}

\begin{document}

\title{Coherent pulsed excitation of degenerate multistate systems: Exact
analytic solutions}
\author{E.S. Kyoseva}
\affiliation{Department of Physics, Sofia University, James Bourchier 5 blvd., 1164
Sofia, Bulgaria}
\affiliation{Fachbereich Physik der Universit\"{a}t, 67653 Kaiserslautern, Germany}
\author{N.V. Vitanov}
\affiliation{Department of Physics, Sofia University, James Bourchier 5 blvd., 1164
Sofia, Bulgaria}
\affiliation{Institute of Solid State Physics, Bulgarian Academy of Sciences,
Tsarigradsko chauss\'{e}e 72, 1784 Sofia, Bulgaria}
\date{\today }

\begin{abstract}
We show that the solution of a multistate system composed of $N$ degenerate
lower (ground) states and one upper (excited) state can be reduced by using
the Morris-Shore transformation to the solution of a two-state system
involving only the excited state and a (bright) superposition of ground
states. In addition, there are $N-1$ dark states composed of ground states.
We use this decomposition to derive analytical solutions for degenerate
extensions of the most popular exactly soluble models: the resonance
solution, the Rabi, Landau-Zener, Rosen-Zener, Allen-Eberly and
Demkov-Kunike models. We suggest various applications of the multistate
solutions, for example, as tools for creating multistate coherent
superpositions by generalized resonant $\pi $-pulses. We show that such
generalized $\pi $-pulses can occur even when the upper state is far off
resonance, at specific detunings, which makes it possible to operate in the
degenerate ground-state manifold without populating the (possibly lossy)
upper state, even transiently.
\end{abstract}

\maketitle

\section{Introduction}

The problem of a two-state quantum system driven by a time-dependent pulsed
external field plays a central role in quantum physics \cite{Shore}. First
of all, this problem is interesting by itself both physically and
mathematically: physically, because the two-state system is the simplest
nontrivial system with discrete energy states in quantum mechanics;
mathematically, because the Schr\"{o}dinger equation for two states poses
interesting mathematical challenges some of which are exactly soluble.
Furthermore, already in the two-state case, important nonclassical phenomena
occur, for instance, the famous Rabi oscillations, which often serve as a
test for quantum behavior, and also provide a powerful tool for coherent
control of quantum dynamics, e.g. by $\pi $ pulses. Finally, in almost all
cases (except for a few exactly soluble), the behavior of a multistate
quantum system can only be understood by reduction to one or more effective
two-state systems, e.g., by adiabatic elimination of weakly coupled states
or by using some intrinsic symmetries.

Besides the well-known solution for exact resonance, there exist several
exactly soluble two-state models, the most widely used being the Rabi \cite%
{Rabi}, Landau-Zener \cite{LZ}, Rosen-Zener \cite{RZ}, Allen-Eberly \cite{AE}%
, Bambini-Berman \cite{BB}, Demkov-Kunike \cite{DK}, Carroll-Hioe \cite{CH},
Demkov \cite{Demkov} and Nikitin \cite{Nikitin} models. All these models
provide the transition probability between two \emph{nondegenerate} states.

\begin{figure}[tbph]
\includegraphics[width=60mm]{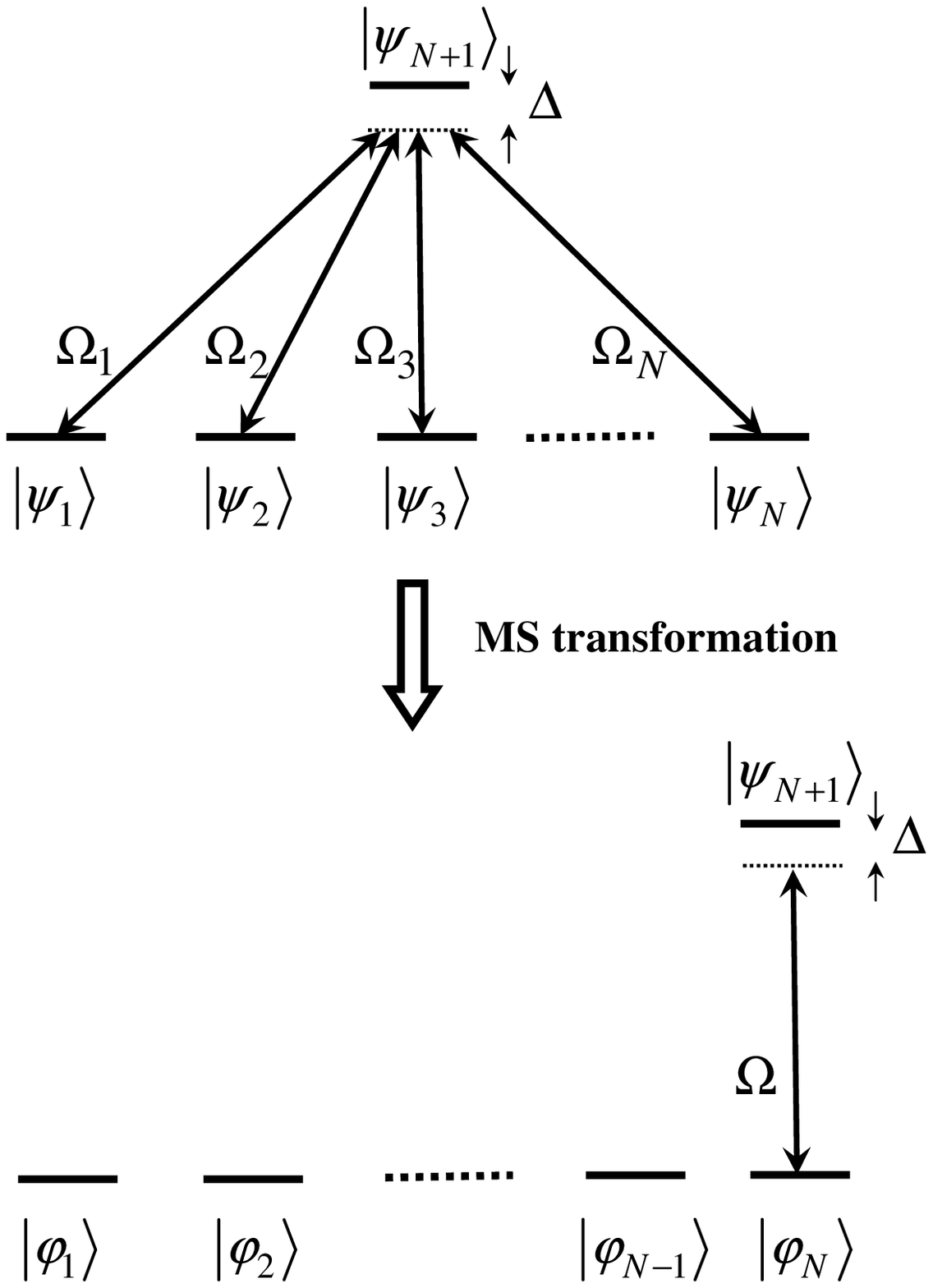}
\caption{Top: The system studied in this paper. $N$ degenerate (in RWA\
sense) states $\left\vert \protect\psi _{1}\right\rangle ,\left\vert \protect%
\psi _{2}\right\rangle ,\ldots ,\left\vert \protect\psi _{N}\right\rangle $
are coupled simultaneously to an upper state $\left\vert \protect\psi %
_{N+1}\right\rangle $, possibly off single-photon resonance by a detuning $%
\Delta (t)$, with Rabi frequencies $\Omega _{n}(t)$ ($n=1,2,\ldots ,N$).
Bottom: The same system in the Morris-Shore basis. There are $N-1$ uncoupled
dark states $\left\vert \protect\varphi _{1}\right\rangle ,\left\vert
\protect\varphi _{2}\right\rangle ,\ldots ,\left\vert \protect\varphi %
_{N-1}\right\rangle $, and a pair of coupled states, a bright state $%
\left\vert \protect\varphi _{N}\right\rangle $ and the upper state $%
\left\vert \protect\psi _{N+1}\right\rangle $, with the same detuning $%
\Delta (t)$ as in the original basis and a coupling given by the rms Rabi
frequency $\Omega (t)$, Eq. (\protect\ref{OmegaRMS}). }%
\label{Fig-system}
\end{figure}

In the present paper, we present the extensions of these exactly soluble
models to the case when one of the states is replaced by $N$ degenerate
states, as displayed in Fig. \ref{Fig-system}. By using the Morris-Shore
(MS) transformation \cite{Morris-Shore} we show that the ($N+1$)-state
problem can be reduced to an effective two-state problem involving a bright
state and the upper, nondegenerate state. If known, the propagator for this
subsystem can be used to find the solution for the full ($N+1$)-state
system. Such analytic solutions can be very useful in designing general
unitary transformations within the $N$-state degenerate manifold, which can
be viewed as a \emph{qunit} for quantum information processing \cite{QI}. We
point out that the same system for $N=3$ has been considered by Unanyan et
al \cite{tripod} and by Kis and Stenholm \cite{Kis} for general $N$, who
have derived the adiabatic solution for pulses generally delayed in time;
these schemes extend the well-known technique of stimulated Raman adiabatic
passage (STIRAP) (see \cite{STIRAP} for reviews). Here we derive several
\emph{exact }analytic solutions for pulses \emph{coincident} in time. This
work can therefore be considered as an extension to arbitrary $N$ of an
earlier paper \cite{Vitanov98}, which treated the case $N=2$.

This paper is organised as follows. In Sec. \ref{Sec-definition} we describe
the system and define the problem. In Sec. \ref{Sec-solution} we introduce
the MS basis and derive the $\left( N+1\right) $-state propagator in terms
of the (presumably known) two-state propagator. In Sec. \ref{Sec-examples}
we use this solution to identify various interesting types of population
evolutions. In Sec. \ref{Sec-applications} we use the analytic solutions for
exact resonance and the Rosen-Zener model to propose several applications,
for example, creation of maximally coherent superpositions and qunit
rotation. In Sec. \ref{Sec-LZ} we discuss some aspects of the multistate
Landau-Zener and Demkov-Kunike models. Finally, Sec. \ref{Sec-conclusions}
provides a summary of the results.

\section{Definition of the problem\label{Sec-definition}}

\subsection{System Hamiltonian}

We consider an $\left( N+1\right) $-state system with $N$ degenerate lower
(ground) states $\left\vert \psi _{n}\right\rangle $ $\left(
n=1,2,...,N\right) $ and one upper (excited) state $\left\vert \psi
_{N+1}\right\rangle $, as depicted in Fig. \ref{Fig-system}. The $N$ lower
states are coupled via the upper state with pulsed interactions, each pair
of which are on two-photon resonance (Fig. \ref{Fig-system}). The upper
state $\left\vert \psi _{N+1}\right\rangle $ may be off single-photon
resonance by some detuning $\Delta (t)$ that, however, must be the same for
all fields. In the usual rotating-wave approximation (RWA) the Schr\"{o}%
dinger equation of the system reads \cite{Shore}%
\begin{equation}
i\hbar \frac{d}{dt}\mathbf{C}(t)=\mathsf{H}(t)\mathbf{C}(t),  \label{SEq}
\end{equation}%
where the elements of the $\left( N+1\right) $-dimensional vector $\mathbf{C}%
(t)$ are the probability amplitudes of the states and the Hamiltonian is
given by%
\begin{equation}
\mathsf{H}(t)=\ \frac{\hbar }{2}%
\begin{bmatrix}
0 & 0 & \cdots  & 0 & \Omega _{1}\left( t\right)  \\
0 & 0 & \cdots  & 0 & \Omega _{2}\left( t\right)  \\
\vdots  & \vdots  & \ddots  & \vdots  & \vdots  \\
0 & 0 & \cdots  & 0 & \Omega _{N}\left( t\right)  \\
\Omega _{1}\left( t\right)  & \Omega _{2}\left( t\right)  & \cdots  & \Omega
_{N}\left( t\right)  & 2\Delta \left( t\right)
\end{bmatrix}%
.  \label{H}
\end{equation}%
For the sake of simplicity the Rabi frequences of the couplings between the
ground states and the excited state $\Omega _{1}\left( t\right) ,...,\Omega
_{N}\left( t\right) $ are assumed real and positive as the populations do
not depend on their signs. The phases of the couplings can easily be
incorporated in the description and they can be used to control the inner
phases of the created superposition states. Furthermore, the Rabi
frequencies are assumed to be pulse-shaped functions with the same time
dependence $f(t)$, but possibly with different magnitudes,%
\begin{equation}
\Omega _{n}\left( t\right) =\chi _{n}f(t)\quad \left( n=1,2,...,N\right) ,
\label{equal (t)}
\end{equation}%
and hence different pulse areas,%
\begin{equation}
A_{n}=\int_{-\infty }^{\infty }\Omega _{n}(t)dt=\chi _{n}\int_{-\infty
}^{\infty }f(t)dt\quad \left( n=1,2,...,N\right) .  \label{area_n}
\end{equation}

\subsection{Physical implementations}

\begin{figure}[tbph]
\includegraphics[width=80mm]{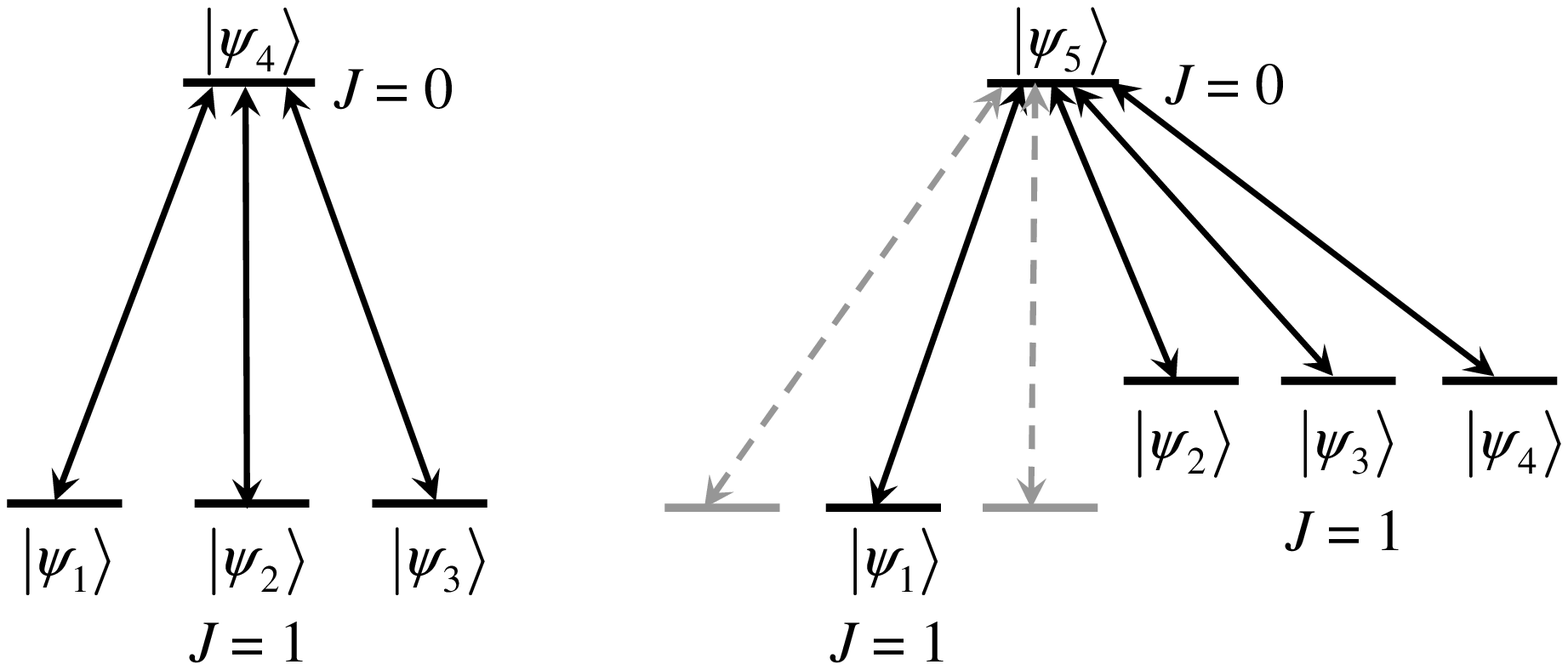}
\caption{Examples of physical implementations of the linkage pattern of $N$
degenerate ground states coupled via one upper state, considered in the
present paper. Left: $N=3$ degenerate states. Right: $N=4$ degenerate (in
the RWA sense) states (dashed arrows indicate two additional possible
linkages).}
\label{Fig-implementations}
\end{figure}

The linkage pattern described by the Hamiltonian (\ref{H}) can be
implemented experimentally in laser excitation of atoms or molecules. For
example, the $N=3$ case is readily implemented in the $J=1\leftrightarrow
J=0 $ system coupled by three laser fields with right circular, left
circular and linear polarizations, as shown in Fig. \ref{Fig-implementations}
(left). These coupling fields can be produced from the same laser by
standard optical tools (beam splitters, polarizers, etc.), which greatly
facilitates implementation. Moreover, the use of pulses derived from the
same laser ensures automatically the two-photon resonance conditions and the
condition (\ref{equal (t)}) for the same temporal profile of all pulses.

The cases of $N=4-6$ can be realized by adding an additional $J=1$ level to
the coupling scheme and appropriately polarized laser pulses, as shown in
the right frame of Fig. \ref{Fig-implementations}.

\section{General solution\label{Sec-solution}}

\subsection{Morris-Shore (dark-bright) basis}

The Hamiltonian (\ref{H}) has $N-1$ zero eigenvalues and two nonzero ones,
\begin{subequations}
\begin{eqnarray}
\lambda _{n} &=&0\quad (n=1,\ldots ,N-1), \\
~\lambda _{\pm }(t) &=&\frac{1}{2}\left[ \Delta \pm \sqrt{\Delta ^{2}+\Omega
^{2}(t)}\right] ,
\end{eqnarray}%
where
\end{subequations}
\begin{equation}
\Omega (t)=\sqrt{\overset{N}{\underset{n=1}{\sum }}\Omega _{n}^{2}(t)}\equiv
\chi f(t)  \label{OmegaRMS}
\end{equation}%
is the root-mean-square (rms) Rabi frequency, where%
\begin{equation}
\chi =\sqrt{\overset{N}{\underset{n=1}{\sum }}\chi _{n}^{2}}.  \label{chi}
\end{equation}%
The set of orthonormalized eigenstates $\left\vert \varphi _{n}\right\rangle
$ $(n=1,2,\ldots ,N-1)$ corresponding to the zero eigenvalues can be chosen
as
\begin{subequations}
\label{dark-states}
\begin{eqnarray}
\left\vert \varphi _{1}\right\rangle &=&\frac{1}{X_{2}}\left[ \chi
_{2},-\chi _{1},0,0,\cdots ,0\right] ^{T},  \label{xi1} \\
\left\vert \varphi _{2}\right\rangle &=&\frac{1}{X_{2}X_{3}}\left[ \chi
_{1}\chi _{3},\chi _{2}\chi _{3},-X_{2}^{2},0,\cdots ,0\right] ^{T},
\label{xi2} \\
\left\vert \varphi _{3}\right\rangle &=&\frac{1}{X_{3}X_{4}}\left[ \chi
_{1}\chi _{4},\chi _{2}\chi _{4},\chi _{3}\chi _{4},-X_{3}^{2},0,\cdots ,0%
\right] ^{T},  \label{xi3} \\
&&\cdots  \notag \\
\left\vert \varphi _{N-1}\right\rangle &=&\frac{1}{X_{N-1}X_{N}}\left[ \chi
_{1}\chi _{N},\chi _{2}\chi _{N},\cdots ,-X_{N-1}^{2},0\right] ^{T},
\label{xiN-1}
\end{eqnarray}%
where
\end{subequations}
\begin{equation}
X_{n}=\sqrt{\overset{n}{\underset{k=1}{\sum }}\chi _{k}^{2}}\quad \left(
n=2,3,...,N\right) .  \label{Xn}
\end{equation}%
These eigenstates are dark states, i.e. they do not involve the excited
state $\left\vert \psi _{N+1}\right\rangle $ and, as we shall see, are
uncoupled from $\left\vert \psi _{N+1}\right\rangle $. All dark states are
time-independent. We emphasize that the choice (\ref{dark-states}) of dark
states is not unique because any superposition of dark states is a dark
state too; hence their choice is a matter of convenience.

The Hilbert space is decomposed into two subspaces: an $(N-1)$-dimensional
dark subspace comprising the dark states (\ref{dark-states}) and a
two-dimensional subspace orthogonal to the dark subspace. It is convenient
to use the Morris-Shore (MS) basis \cite{Morris-Shore}, which, in addition
to the dark states, includes the excited state\ $\left\vert \psi
_{N+1}\right\rangle \equiv \left\vert \varphi _{N+1}\right\rangle $ and a
bright ground state $\left\vert \varphi _{N}\right\rangle $. The latter does
not have a component of the excited state and is orthogonal to the dark
states; these conditions determine it completely (up to an unimportant
global phase),%
\begin{equation}
\left\vert \varphi _{N}\right\rangle =\frac{1}{X_{N}}\left[ \chi _{1},\chi
_{2},\cdots ,\chi _{N},0\right] ^{T}.  \label{bright state}
\end{equation}%
We point out that the Morris-Shore basis is \emph{not} the adiabatic basis
because only the dark states are eigenstates of the Hamiltonian, but $%
\left\vert \varphi _{N}\right\rangle $ and $\left\vert \varphi
_{N+1}\right\rangle $ are not.

In the new, still stationary basis $\left\{ \left\vert \varphi
_{n}\right\rangle \right\} _{n=1,2,...,N+1}$, the Schr\"{o}dinger equation
reads
\begin{equation}
i\hbar \frac{d}{dt}\mathbf{B}(t)=\widetilde{\mathsf{H}}(t)\mathbf{B}(t),
\label{SEq-MS}
\end{equation}%
where the original amplitudes $\mathbf{C}(t)$ are connected to the MS
amplitudes $\mathbf{B}(t)$ by the time-independent unitary matrix $\mathsf{W}
$ composed by the basis vectors $\left\vert \varphi _{n}\right\rangle $,
\begin{equation}
\mathsf{W}=\left[ \left\vert \varphi _{1}\right\rangle ,\left\vert \varphi
_{2}\right\rangle ,~\ldots ,~\left\vert \varphi _{N+1}\right\rangle \right] ,
\label{W}
\end{equation}%
according to
\begin{equation}
\mathbf{C}(t)=\mathsf{W}\mathbf{B}(t).  \label{C=WA}
\end{equation}%
The transformed Hamiltonian reads $\widetilde{\mathsf{H}}(t)=\mathsf{W}%
^{\dag }\mathsf{H}(t)\mathsf{W}$, or explicitly,%
\begin{equation}
\widetilde{\mathsf{H}}(t)=\frac{\hbar }{2}%
\begin{bmatrix}
0 & 0 & \cdots  & 0 & 0 & 0 \\
0 & 0 & \cdots  & 0 & 0 & 0 \\
\vdots  & \vdots  & \ddots  & \vdots  & \vdots  & \vdots  \\
0 & 0 & \cdots  & 0 & 0 & 0 \\
0 & 0 & \cdots  & 0 & 0 & \Omega (t) \\
0 & 0 & \cdots  & 0 & \Omega (t) & 2\Delta (t)%
\end{bmatrix}%
.  \label{H-MS}
\end{equation}

We point out that the Hamiltonian of Eq. (\ref{H}) is a special case of the
most general Hamiltonian for which the MS\ transformation \cite{Morris-Shore}
applies and which includes $N$ degenerate lower states and $M$ degenerate
upper states. Hamiltonians of the same type as (\ref{H}) and related
transformations leading to Eq. (\ref{H-MS}), have appeared in the literature
also before the paper by Morris and Shore \cite{Morris-Shore}, mostly in simplified
versions of constant and equal interactions (see e.g. \cite%
{Stenholm} and references therein).

\subsection{Solution in the Morris-Shore basis}

As evident from the first $N-1$ zero rows of $\widetilde{\mathsf{H}}$ the
dark states are decoupled from states $\left\vert \varphi _{N}\right\rangle $
and $\left\vert \varphi _{N+1}\right\rangle $ and the dark-state amplitudes
remain unchanged, $B_{n}(t)=const$ ($n=1,2,\ldots ,N-1$). Thus the $\left(
N+1\right) $-state problem reduces to a two-state one involving $\left\vert
\varphi _{N}\right\rangle $ and $\left\vert \varphi _{N+1}\right\rangle $,%
\begin{equation}
i\frac{d}{dt}\left[
\begin{array}{c}
B_{N} \\
B_{N+1}%
\end{array}%
\right] =\frac{1}{2}%
\begin{bmatrix}
0 & \Omega \\
\Omega & 2\Delta%
\end{bmatrix}%
\left[
\begin{array}{c}
B_{N} \\
B_{N+1}%
\end{array}%
\right] .  \label{2 state problem}
\end{equation}%
The propagator for this two-state system, defined by%
\begin{equation}
\left[
\begin{array}{c}
B_{N}\left( +\infty \right) \\
B_{N+1}\left( +\infty \right)%
\end{array}%
\right] =\mathsf{U}_{MS}^{\left( 2\right) }\left[
\begin{array}{c}
B_{N}\left( -\infty \right) \\
B_{N+1}\left( -\infty \right)%
\end{array}%
\right] ,  \label{A=UA}
\end{equation}%
is unitary and can be expressed in terms of the Cayley-Klein parameters as%
\begin{equation}
\mathsf{U}_{MS}^{\left( 2\right) }=%
\begin{bmatrix}
a & b \\
-b^{\ast } & a^{\ast }%
\end{bmatrix}%
,  \label{U2}
\end{equation}%
with $\left\vert b\right\vert ^{2}=1-\left\vert a\right\vert ^{2}$. Then the
transition matrix for the $\left( N+1\right) $-state system in the MS basis
reads
\begin{equation}
\mathsf{U}_{MS}^{\left( N+1\right) }=%
\begin{bmatrix}
1 & 0 & \cdots & 0 & 0 & 0 \\
0 & 1 & \cdots & 0 & 0 & 0 \\
\vdots & \vdots & \ddots & \vdots & \vdots & \vdots \\
0 & 0 & \cdots & 1 & 0 & 0 \\
0 & 0 & \cdots & 0 & a & b \\
0 & 0 & \cdots & 0 & -b^{\ast } & a^{\ast }%
\end{bmatrix}%
.  \label{U-MS}
\end{equation}

\subsection{The solution in the original basis}

We can find the transition matrix in the original, diabatic basis
by using the transformation%
\begin{equation}
\mathsf{U}^{\left( N+1\right) }(\infty ,-\infty )=\mathsf{WU}_{MS}^{\left(
N+1\right) }(\infty ,-\infty )\mathsf{W}^{\dag },  \label{U=WUW}
\end{equation}%
or explicitly,%
\begin{widetext}
\begin{equation}
\mathsf{U}^{\left( N+1\right) }=%
\begin{bmatrix}
1+\left( a-1\right) \frac{\chi _{1}^{2}}{\chi ^{2}} & \left( a-1\right)
\frac{\chi _{1}\chi _{2}}{\chi ^{2}} & \left( a-1\right) \frac{\chi
_{1}\chi _{3}}{\chi ^{2}} & \cdots  & \left( a-1\right) \frac{\chi
_{1}\chi _{N}}{\chi ^{2}} & b\frac{\chi _{1}}{\chi } \\
\left( a-1\right) \frac{\chi _{1}\chi _{2}}{\chi ^{2}} & 1+\left(
a-1\right) \frac{\chi _{2}^{2}}{\chi ^{2}} & \left( a-1\right) \frac{%
\chi _{2}\chi _{3}}{\chi ^{2}} & \cdots  & \left( a-1\right) \frac{%
\chi _{2}\chi _{N}}{\chi ^{2}} & b\frac{\chi _{2}}{\chi } \\
\left( a-1\right) \frac{\chi _{1}\chi _{3}}{\chi ^{2}} & \left(
a-1\right) \frac{\chi _{2}\chi _{3}}{\chi ^{2}} & 1+\left( a-1\right)
\frac{\chi _{3}^{2}}{\chi ^{2}} & \cdots  & \left( a-1\right) \frac{%
\chi _{3}\chi _{N}}{\chi ^{2}} & b\frac{\chi _{3}}{\chi } \\
\vdots  & \vdots  & \vdots  & \ddots  & \vdots  & \vdots  \\
\left( a-1\right) \frac{\chi _{1}\chi _{N}}{\chi ^{2}} & \left(
a-1\right) \frac{\chi _{2}\chi _{N}}{\chi ^{2}} & \left( a-1\right)
\frac{\chi _{3}\chi _{N}}{\chi ^{2}} & \cdots  & 1+\left( a-1\right)
\frac{\chi _{N}^{2}}{\chi ^{2}} & b\frac{\chi _{N}}{\chi } \\
-b^*\frac{\chi _{1}}{\chi } & -b^*\frac{\chi _{2}}{\chi } & -b^*\frac{\chi
_{3}}{\chi } & \cdots  & -b^*\frac{\chi _{N}}{\chi } & a^*%
\end{bmatrix}%
.  \label{U}
\end{equation}%
\end{widetext}The $i$th column of this matrix provides the probability
amplitudes for initial conditions
\begin{subequations}
\label{initial condition k}
\begin{eqnarray}
C_{i}(-\infty ) &=&1, \\
C_{n}(-\infty ) &=&0\quad (n\neq i).
\end{eqnarray}%
The initial state $\left\vert \psi _{i}\right\rangle $ can be one of the
degenerate states or the upper state. This general unitary matrix and
combinations of such matrices can be used to design techniques for general
or special qunit rotations.

As evident from Eq. (\ref{U}) for finding the populations for the initial
condition (\ref{initial condition k}) it is sufficient to know only the parameter $a=\left[
U_{MS}^{(2)}(\infty ,-\infty )\right] _{11}$ because $\left\vert
b\right\vert ^{2}=1-\left\vert a\right\vert ^{2}$ \cite{Vitanov98}.
For the sake of simplicity, in the present paper we are interested only in
cases when the system starts in a single state and below we shall
concentrate on the values of the parameter $a$. In the more general case
when the system starts in a coherent superposition of states, Eq. (\ref{U})
can be used again to derive the solution; then the other Cayley-Klein
parameter $b$ is also needed.

\section{Types of population distribution\label{Sec-examples}}

We identify two types of initial conditions: when the system starts in one
of the degenerate states $\left\vert \psi _{i}\right\rangle $ or in the
excited state $\left\vert \psi _{N+1}\right\rangle $, which we shall
consider separately.

\subsection{System initially in a ground state}

\subsubsection{General case}

When the system is initially in the ground state $\left\vert \psi
_{i}\right\rangle $, Eq. (\ref{initial condition k}), we find from the $i$th
column of the propagator (\ref{U}) that the populations in the end of the
evolution are
\end{subequations}
\begin{subequations}
\label{populations psi_i}
\begin{eqnarray}
P_{i} &=&\left\vert 1+\left( a-1\right) \frac{\chi _{i}^{2}}{\chi ^{2}}%
\right\vert ^{2}, \\
P_{n} &=&\left\vert a-1\right\vert ^{2}\frac{\chi _{i}^{2}\chi _{n}^{2}}{%
\chi ^{4}}\quad (n\neq i,N+1), \\
P_{N+1} &=&\left( 1-\left\vert a\right\vert ^{2}\right) \frac{\chi _{i}^{2}}{%
\chi ^{2}}.
\end{eqnarray}%
Therefore the ratio of the populations of any two degenerate states,
different from the initial state $\left\vert \psi _{i}\right\rangle $, reads
\end{subequations}
\begin{equation}
\frac{P_{m}}{P_{n}}=\frac{\chi _{m}^{2}}{\chi _{n}^{2}}\quad (m,n\neq i,N+1).
\label{Pm/Pn}
\end{equation}%
Hence these population ratios do not depend on the interaction details but
only on the ratios of the corresponding peak Rabi frequencies.

For equal Rabi frequencies,
\begin{equation}
\chi _{1}=\chi _{2}=\cdots =\chi _{N},  \label{Omega equal}
\end{equation}%
Eqs. (\ref{populations psi_i}) reduce to
\begin{subequations}
\label{P for Omega equal}
\begin{eqnarray}
P_{i} &=&\left\vert 1+\frac{a-1}{N}\right\vert ^{2},
\label{Pn for Omega equal} \\
P_{n} &=&\frac{\left\vert a-1\right\vert ^{2}}{N^{2}}\quad (n\neq i,N+1), \\
P_{N+1} &=&\frac{1-\left\vert a\right\vert ^{2}}{N}.
\label{P0 for Omega equal}
\end{eqnarray}%
Thus the populations of all ground states except the initial state $%
\left\vert \psi _{i}\right\rangle $ are equal.

\subsubsection{Special values of $a$}

Several values of the propagator parameter $a$ are especially interesting.

For $a=0$, which indicates complete population transfer (CPT) in the MS
two-state system, Eq. (\ref{populations psi_i}) gives
\end{subequations}
\begin{subequations}
\label{P for a=0}
\begin{eqnarray}
P_{i} &=&\left\vert 1-\frac{\chi _{i}^{2}}{\chi ^{2}}\right\vert ^{2}, \\
P_{n} &=&\frac{\chi _{n}^{2}\chi _{i}^{2}}{\chi ^{4}}\quad (n\neq i,N+1),
\label{P0 for a=0} \\
P_{N+1} &=&\frac{\chi _{i}^{2}}{\chi ^{2}}.
\end{eqnarray}%
\newline

For $a=1$, which corresponds to complete population return (CPR) in the MS
two-state system, we obtain
\end{subequations}
\begin{subequations}
\label{P for a=1}
\begin{eqnarray}
P_{i} &=&1,  \label{Pn for a=1} \\
P_{n} &=&0\quad (n\neq i,N+1), \\
P_{N+1} &=&0.  \label{P0 for a=1}
\end{eqnarray}

For $a=-1$, which again corresponds to CPR in the MS two-state system, but
with a sign flip in the amplitude, we find
\end{subequations}
\begin{subequations}
\label{P for a=-1}
\begin{eqnarray}
P_{i} &=&\left( 1-2\frac{\chi _{i}^{2}}{\chi ^{2}}\right) ^{2},
\label{Pk for a=-1} \\
P_{n} &=&\frac{4\chi _{n}^{2}\chi _{i}^{2}}{\chi ^{4}}\quad (n\neq i,N+1), \\
P_{N+1} &=&0.  \label{P0 for a=-1}
\end{eqnarray}%
It is important to note that although both cases $a=1$ and $a=-1$ lead to
CPR in the MS\ two-state system, they produce very different population
distributions in the full $\left( N+1\right) $-state system. The case $a=1$
leads to a trivial result (CPR in the full system), whereas the case $a=-1$
is very interesting because it leads to a population redistribution amongst
the ground states with zero population in the upper state; hence this case
deserves a special attention.

\subsubsection{The case $a=-1$}

The case of $a=-1$ is particularly important because it allows to create a
coherent superposition of all ground states, with no population in the upper
state.

All ground-state populations in this superposition will be equal,
\end{subequations}
\begin{subequations}
\label{P equal}
\begin{eqnarray}
&&P_{1}=P_{2}=\cdots =P_{N}=\frac{1}{N}, \\
&&P_{N+1}=0.
\end{eqnarray}%
if
\end{subequations}
\begin{subequations}
\label{Case I}
\begin{eqnarray}
\chi _{i}=\left( \sqrt{N}\pm 1\right) \chi _{0}, && \\
\chi _{n}=\chi _{0}\quad (n\neq i), &&
\end{eqnarray}%
where%
\begin{equation}
\chi _{0}=\frac{\chi }{\sqrt{2\left( N\pm \sqrt{N}\right) }}.
\end{equation}%
This result does not depend on other interaction details (pulse shape, pulse
area, detuning) as long as $a=-1$. For example, for $N=4$ degenerate states,
equal populations are obtained when $\chi _{i}=\chi _{n}$ or $\chi
_{i}=3\chi _{n}$. We shall discuss later how the condition $a=-1$ can be
obtained for several analytically soluble models.

Another important particular case is when the initial-state population $%
P_{i} $ vanishes in the end. This occurs for
\end{subequations}
\begin{equation}
\chi _{i}^{2}=\sum_{n\neq i}\chi _{n}^{2}.  \label{condition equal Pn, Pk=0}
\end{equation}%
For example, an equal superposition of all lower sublevels except $%
\left\vert \psi _{i}\right\rangle $,
\begin{subequations}
\label{P equal w/o Pi}
\begin{eqnarray}
P_{i} &=&P_{N+1}=0, \\
P_{n} &=&\frac{1}{N-1}\quad (n\neq i,N+1),
\end{eqnarray}%
is created for
\end{subequations}
\begin{subequations}
\label{Case II}
\begin{eqnarray}
\chi _{i} &=&\chi _{0}\sqrt{N-1}, \\
\chi _{n} &=&\chi _{0}\quad (n\neq i),
\end{eqnarray}%
where%
\begin{equation}
\chi _{0}=\frac{\chi }{\sqrt{2\left( N-1\right) }}.
\end{equation}

\subsection{System initially in the upper state\label{Sec-upper}}

If the system is initially in the excited state $\left\vert \psi
_{N+1}\right\rangle $, at the end of the evolution the populations will be
\end{subequations}
\begin{subequations}
\label{populations psi0}
\begin{eqnarray}
&&P_{n}=\left( 1-\left\vert a\right\vert ^{2}\right) \frac{\chi _{n}^{2}}{%
\chi ^{2}}\quad \left( n=1,2,\ldots ,N\right) , \\
&&P_{N+1}=\left\vert a\right\vert ^{2}.
\end{eqnarray}%
For $a=\pm 1$ at the end of the evolution the system undergoes CPR, as in
the MS two-state system. For $a=0$ (CPT in the MS two-state system) the
whole population will be in the ground states leaving the excited state
empty, $P_{N+1}=0$. If all the couplings are equal, Eq. (\ref{Omega equal}),
the ground states will have equal populations,
\end{subequations}
\begin{equation}
P_{n}=\frac{1}{N}\quad \left( n=1,2,\ldots ,N\right) .
\label{Pequal from Pe}
\end{equation}

\subsection{Discussion\label{Sec-discussion}}

In this section we discussed some general features of the population
redistribution in the $(N+1)$-state system. There are three particularly
interesing results.

\emph{First}, the ratios of the populations of the degenerate states
(except the one populated initially) depend only on the ratios of the
corresponding Rabi frequencies; hence they can be controlled by
changing the corresponding laser intensities alone. The populations values, though,
depend on the other interaction details. Moreover, it can easily be
seen that the relative phases of the degenerate states can be controlled by
the relative laser phases.

\emph{Second}, it is possible to create an equal superposition of all ground
states, with zero population in the upper state. This is possible when the
system starts in a ground state: then condition (\ref{Case I}) is required,
along with the CPR\ condition $a=-1$. Alternatively, an equal superposition
can be created when the system starts in the upper state: then condition (%
\ref{Omega equal}) is required, along with the CPT condition $a=0$. Equal
superpositions are important in some applications because they are states
with maximal coherence (since the population inversions vanish).

\emph{Third}, it is possible, starting from a ground state, to create a
superposition of all other ground states, whereas the initial ground state
and the excited state are left unpopulated. This requires $a=-1$ and
condition (\ref{condition equal Pn, Pk=0}). This case has interesting
physical implications, which will be discussed in the next section.

\section{Applications to exactly soluble models\label{Sec-applications}}

\subsection{Multistate analytical solutions}

\begin{table}[t]
\begin{tabular}{|l|}
\colrule Model \\
\colrule Resonance \\
$\Omega (t)=\chi f(t),\quad \Delta (t)=0$ \\
$a=\cos \frac{1}{2}A$ \\
\colrule Rabi \\
$\Omega (t)=\chi $ $(\left\vert t\right\vert \leqq T),\quad \Delta
(t)=\Delta _{0}$ \\
$a=\cos \left( T\sqrt{\chi ^{2}+\Delta ^{2}}\right) -i\dfrac{\Delta }{\sqrt{%
\Omega ^{2}+\Delta ^{2}}}\sin \left( T\sqrt{\chi ^{2}+\Delta ^{2}}\right) $
\\
\colrule Landau-Zener \\
$\Omega (t)=\chi ,\quad \Delta (t)=Ct$ \\
$a=\exp \left[ -\pi \chi ^{2}/4C\right] $ \\
\colrule Rosen-Zener \\
$\Omega (t)=\chi $sech$(t/T),\quad \Delta (t)=\Delta _{0}$ \\
$a=\dfrac{\Gamma ^{2}\left( \frac{1}{2}+i\delta \right) }{\Gamma \left(
\frac{1}{2}+\alpha +i\delta \right) \Gamma \left( \frac{1}{2}-\alpha
+i\delta \right) }$ \\
\colrule Allen-Eberly \\
$\Omega (t)=\chi $sech$(t/T),\quad \Delta (t)=B\tanh (t/T)$ \\
$a=\dfrac{\cos \left( \pi \sqrt{\alpha ^{2}-\beta^{2}}\right) }{\cosh
\left( \pi \beta\right) }$ \\
\colrule Demkov-Kunike \\
$\Omega (t)=\chi $sech$(t/T),\quad \Delta (t)=\Delta _{0}+B\tanh (t/T)$ \\
$a=\dfrac{\Gamma \left( \frac{1}{2}+i(\delta +\beta )\right) \Gamma \left(
\frac{1}{2}+i(\delta -\beta )\right) }{\Gamma \left( \frac{1}{2}+\sqrt{%
\alpha ^{2}-\beta ^{2}}+i\delta \right) \Gamma \left( \frac{1}{2}-\sqrt{%
\alpha ^{2}-\beta ^{2}}+i\delta \right) }$ \\
\colrule%
\end{tabular}%
\caption{Values of the Cayley-Klein parameter $a=\left[ U_{MS}^{(2)}(\infty
,-\infty )\right] _{11}$ for several exactly soluble models. Here $\Gamma
(z) $ is the Gamma function and $\protect\alpha =\frac{1}{2}\protect\chi T$,
$\protect\beta =\frac{1}{2}BT$, $\protect\delta =\frac{1}{2}\Delta _{0}T$,
are scaled dimensionless parameters, which are assumed positive without loss
of generality.}
\label{Table-a-exact}%
\end{table}
The values of the propagator parameter $a=\left[ U_{MS}^{(2)}(\infty
,-\infty )\right] _{11}$ for the most popular analytically exactly soluble
models are listed in Table \ref{Table-a-exact}. Equation (\ref{U}), supplied
with these values, provides several exact multistate analytical solutions,
which generalize the respective two-state solutions.

Among these solutions, the resonance case is the simplest and most important
one, which will receive a special attention below. It will be followed by a
detailed discussion of the Rosen-Zener (RZ) model, which can be seen as an
extension of the resonance solution to nonzero detuning for a special pulse
shape (hyperbolic secant). Both the resonance and the RZ model allow for the
parameter $a$ to obtain the important values $0,\pm 1$. The Rabi model can
also be used to illustrate the interesting cases of population distribution
associated with these values of $a$ but its rectangular pulse shape
is less attractive (and also less realistic) than the beautiful sech-shape
of the pulse in the RZ model.

The Landau-Zener (LZ) and Allen-Eberly (AE) models are of level-crossing
type, i.e. the detuning crosses resonance, $\Delta (0)=0$. For these models in the adiabatic
limit the transition probability approaches unity, that is $a\rightarrow 0$.
The parameter $a$ is always nonnegative, i.e. the most interesting value in
the present context, $a=-1$, is unreachable. Nevertheless, because of the
popularity and the importance of the LZ model, and because the present multistate LZ
solution supplements other multistate LZ solutions, we discuss this solution
in detail in Sec. \ref{Sec-LZ}.

The Demkov-Kunike (DK) model is a very versatile model, which combines and
generalizes the RZ and AE models. Indeed, as seen in Table \ref%
{Table-a-exact}, the DK model reduces to the RZ model for $B=0$ and to the
AE model for $\Delta _{0}=0$. For the DK model, the parameter $a$ can be
equal to the most interesting value of $-1$ only when $B=0$, i.e. only in
the RZ limit. Therefore, we shall only consider the RZ model below, and
leave the AE and DK models to readers interested in other aspects of the
analytic multistate solutions presented here.

\subsection{Exact resonance\label{Sec-resonance}}

In the case of exact resonance,
\begin{equation}
\Delta =0,
\end{equation}%
the elements of the evolution matrix for the MS two-state system for any
pulse shape of $\Omega (t)$ are
\begin{subequations}
\label{resonance}
\begin{eqnarray}
a &=&\cos \frac{A}{2},~~ \\
b &=&-i\sin \frac{A}{2},
\end{eqnarray}%
where $A$ is the rms pulse area\ defined as
\end{subequations}
\begin{equation}
A=\int_{-\infty }^{\infty }\Omega (t^{\prime })dt^{\prime }.
\label{pulse area}
\end{equation}

In the important case of $N=3$ we have%
\begin{widetext}
\begin{equation}
\mathsf{U}_{d}^{\left( 4\right) }=%
\begin{bmatrix}
1-2\frac{\chi _{1}^{2}}{\chi ^{2}}\sin ^{2}\frac{1}{4}A & -2\frac{\chi
_{1}\chi _{2}}{\chi ^{2}}\sin ^{2}\frac{1}{4}A & -2\frac{\chi
_{1}\chi _{3}}{\chi ^{2}}\sin ^{2}\frac{1}{4}A & -i\frac{\chi _{1}}{%
\chi }\sin \frac{1}{2}A \\
-2\frac{\chi _{1}\chi _{2}}{\chi ^{2}}\sin ^{2}\frac{1}{4}A & 1-2\frac{%
\chi _{2}^{2}}{\chi ^{2}}\sin ^{2}\frac{1}{4}A & -2\frac{\chi
_{2}\chi _{3}}{\chi ^{2}}\sin ^{2}\frac{1}{4}A & -i\frac{\chi _{2}}{%
\chi }\sin \frac{1}{2}A \\
-2\frac{\chi _{1}\chi _{3}}{\chi ^{2}}\sin ^{2}\frac{1}{4}A & -2\frac{%
\chi _{2}\chi _{3}}{\chi ^{2}}\sin ^{2}\frac{1}{4}A & 1-2\frac{\chi
_{3}^{2}}{\chi ^{2}}\sin ^{2}\frac{1}{4}A & -i\frac{\chi _{3}}{\chi }%
\sin \frac{1}{2}A \\
-i\frac{\chi _{1}}{\chi }\sin \frac{1}{2}A & -i\frac{\chi _{2}}{\chi
}\sin \frac{1}{2}A & -i\frac{\chi _{3}}{\chi }\sin \frac{1}{2}A & \cos
\frac{1}{2}A%
\end{bmatrix}%
.  \label{U4-resonance}
\end{equation}%
\end{widetext}

We have $a=0,\pm 1$ for the following pulse areas,
\begin{subequations}
\label{a resonance}
\begin{eqnarray}
a=0:\quad &&A=\left( 2l+1\right) \pi , \\
a=1:\quad &&A=4l\pi , \\
a=-1:\quad &&A=2\left( 2l+1\right) \pi .
\end{eqnarray}%
where $l=0,1,2,...$.

The pulse areas for the three important cases discussed in Sec. \ref%
{Sec-discussion} are easily calculated.

An equal superposition of all $N$ ground states is created when starting
from the excited state and all individual pulse areas are equal to (see Sec. %
\ref{Sec-upper})
\end{subequations}
\begin{equation}
A_{n}=\frac{\left( 2l+1\right) \pi }{\sqrt{N}}\quad \left( n=1,2,\ldots
,N\right) ,
\end{equation}%
where $l=0,1,2,...$.

An equal superposition of all $N$ ground states is created also when
starting from one ground state $\left\vert \psi _{i}\right\rangle $ and the
pulse areas are [see Eq. (\ref{Case I})]
\begin{subequations}
\begin{eqnarray}
A_{i} &=&\sqrt{2\frac{\sqrt{N}\pm 1}{\sqrt{N}}}\left( 2l+1\right) \pi , \\
A_{n} &=&\sqrt{\frac{2}{N\pm \sqrt{N}}}\left( 2l+1\right) \pi \quad \left(
n\neq i\right) ,
\end{eqnarray}%
where $l=0,1,2,...$

The other interesting case when the system starts in one ground state $%
\left\vert \psi _{i}\right\rangle $ and ends up in an equal superposition of
all other ground states is realised for pulse areas [see Eq. (\ref{Case II}%
)]
\end{subequations}
\begin{subequations}
\begin{eqnarray}
A_{i} &=&\sqrt{2}\left( 2l+1\right) \pi , \\
A_{n} &=&\sqrt{\frac{2}{N-1}}\left( 2l+1\right) \pi \quad \left( n\neq
i\right) ,
\end{eqnarray}%
where $l=0,1,2,...$.

\subsection{Multistate Rosen-Zener model}

Equation (\ref{U}) and the value of the parameter $a$ in Table \ref%
{Table-a-exact} represent the multistate RZ solution in the degenerate
two-level system. It is easy to show that
\end{subequations}
\begin{equation}
\left\vert a\right\vert ^{2}=1-\frac{\sin ^{2}\left( \frac{1}{2}\pi \chi
T\right) }{\cosh ^{2}\left( \frac{1}{2}\pi \Delta _{0}T\right) },
\label{|a|-RZ}
\end{equation}%
where we have used the reflection formula $\Gamma (\frac{1}{2}+z)\Gamma (%
\frac{1}{2}-z)=\pi /\cos \pi z$ \cite{AS}. Hence in this model $\left\vert
a\right\vert =1$ for $\alpha =\frac{1}{2}\chi T=l$ ($l=0,1,2,...$). The
phase of $a$, however, depends on the detuning $\Delta _{0}$ \cite{Vitanov98}%
; we use this to an advantage to select values of $\Delta _{0}$ for which $%
a=-1$. For $\alpha =l$ we find \cite{Vitanov98}
\begin{equation}
a=(-1)^{n}\prod_{k=0}^{n-1}\frac{2l+1-i\Delta _{0}T}{2l+1+i\Delta _{0}T}%
\quad (\alpha =l),  \label{CPT}
\end{equation}%
where the recurrence relation $\Gamma (z+1)=z\Gamma (z)$ \cite{AS} has been
used. Thus, the equation $a=-1$ reduces to an algebraic equation for $\Delta
_{0}$, which has $l$ real solutions \cite{Vitanov98}. The first few values
of $\chi $ and $\Delta _{0}$ for which $a=-1$ are shown in Table \ref%
{Table-a}. As the table shows, $\Delta _{0}=0$ is a solution for odd $\alpha
=\frac{1}{2}\chi T$ but not for even $\alpha $, in agreement with the
conclusions in Sec. \ref{Sec-resonance}. Moreover, the $a=-1$ solutions
do not depend on the number of degenerate states $N$.

\begin{table}[tbp]
\begin{tabular}{|r|rrrrrr|}
\colrule$\chi T$ & $\Delta _{0}T$ &  &  &  &  &  \\
\colrule2 & 0 &  &  &  &  &  \\
4 & $\pm 1.732$ &  &  &  &  &  \\
6 & 0 & $\pm 4.796$ &  &  &  &  \\
8 & $\pm 1.113$ & $\pm 9.207$ &  &  &  &  \\
10 & 0 & $\pm 2.756$ & $\pm 14.913$ &  &  &  \\
12 & $\pm 0.943$ & $\pm 4.936$ & $\pm 21.903$ &  &  &  \\
14 & 0 & $\pm 2.243$ & $\pm 7.595$ & $\pm 30.171$ &  &  \\
16 & $\pm 0.855$ & $\pm 3.916$ & $\pm 10.708$ & $\pm 39.715$ &  &  \\
18 & 0 & $\pm 1.988$ & $\pm 5.907$ & $\pm 14.265$ & $\pm 50.534$ &  \\
20 & $\pm 0.799$ & $\pm 3.418$ & $\pm 8.195$ & $\pm 18.260$ & $\pm 62.627$ &
\\
22 & 0 & $\pm 1.830$ & $\pm 5.098$ & $\pm 10.766$ & $\pm 22.687$ & $\pm
75.993$ \\
24 & $\pm 0.759$ & $\pm 3.113$ & $\pm 7.006$ & $\pm 13.613$ & $\pm 27.545$ &
$\pm 90.634$ \\
26 & 0 & $\pm 1.719$ & $\pm 4.606$ & $\pm 9.130$ & $\pm 16.729$ & $\pm
32.833 $ \\
&  & $\pm 106.549$ &  &  &  &  \\
28 & $\pm 0.728$ & $\pm 2.901$ & $\pm 6.289$ & $\pm 11.461$ & $\pm 20.113$ &
$\pm 38.548$ \\
&  & $\pm 123.736$ &  &  &  &  \\
30 & 0 & $\pm 1.636$ & $\pm 4.268$ & $\pm 8.150$ & $\pm 13.994$ & $\pm
23.760 $ \\
&  & $\pm 44.690$ & $\pm 142.198$ &  &  &  \\
\colrule
\end{tabular}%
\caption{Some approximate solutions of the equation $a(\Delta _{0})=-1$ for
the RZ model, where $a$ is given in Table \protect\ref{Table-a-exact}, for
various even integer values of $\protect\chi T$.}
\label{Table-a}%
\end{table}

In the present context the RZ model is interesting for it shows that one can
create superpositions within the ground-state manifold even when the excited
state is off resonance by a considerable detuning ($\Delta _{0}\gg 1/T$),
for which the transition probability in the MS\ two-state system is
virtually zero, i.e. $\left\vert a\right\vert \approx 1$. This fact allows
us, for specific detunings, to essentially contain the \emph{transient}
dynamics within the ground states; in contrast, in the resonance case the
excited state can get significant transient population, $P_{N+1}(t)=\sin ^{2}%
\frac{1}{2}A(t)$, although it vanishes in the end.

\begin{figure}[tbp]
\includegraphics[width=75mm]{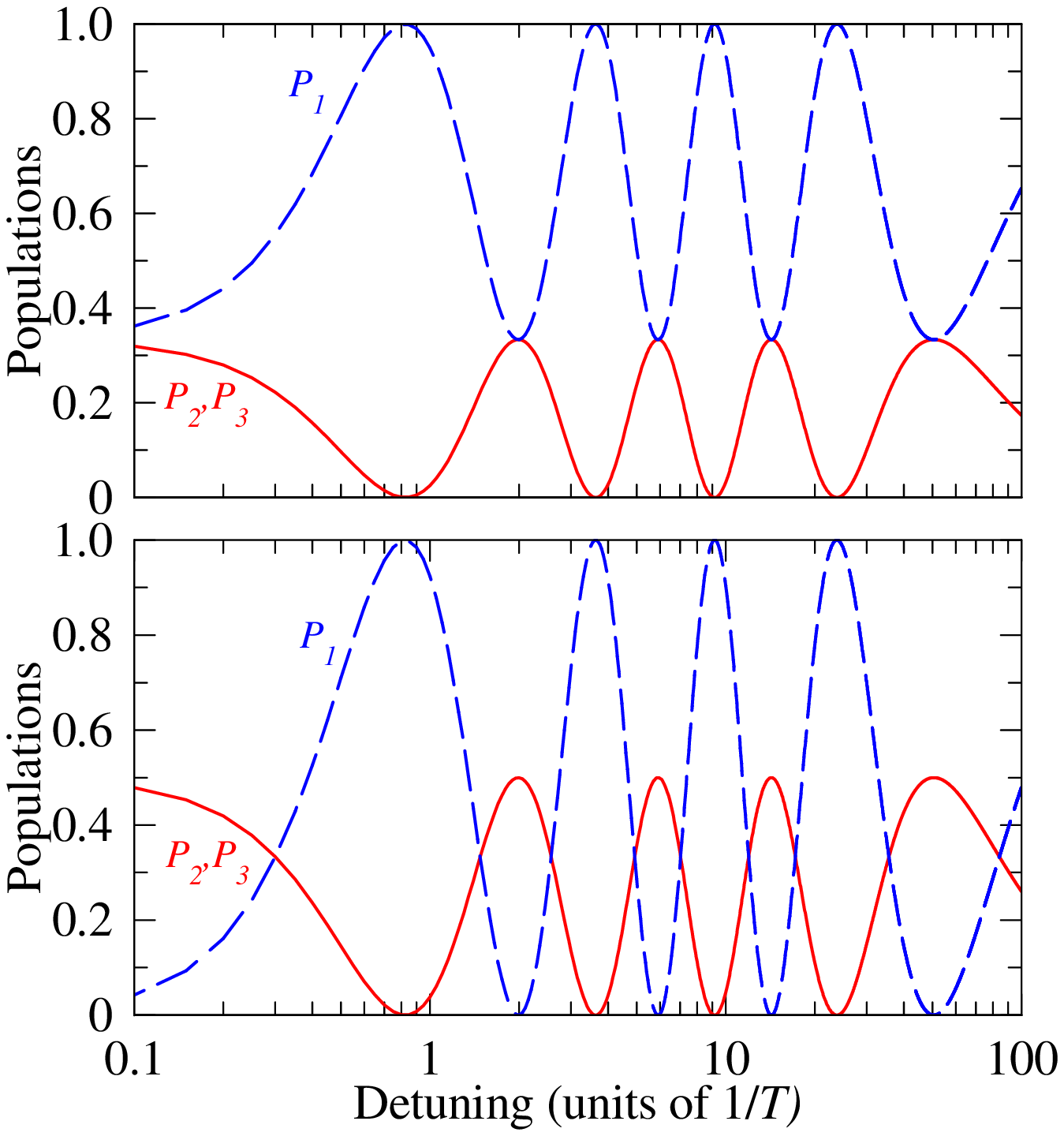}
\caption{(Color online) Populations vs the detuning $\Delta _{0}$ for $N=3$
lower states and $\protect\chi T=18$. The coupling strengths $\protect\chi %
_{n}$ are given by Eqs. (\protect\ref{Case I}) in the upper frame and Eqs. (%
\protect\ref{Case II}) in the lower frame. The system is initially in state $%
\left\vert \protect\psi _{1}\right\rangle $.}
\label{Fig-detuning}
\end{figure}

Figure \ref{Fig-detuning} displays the populations against the detuning $%
\Delta _{0}$ for a hyperbolic-secant pulse with $\chi T=18$ for couplings
chosen to satisfy Eqs. (\ref{Case I}) (upper frame) and (\ref{Case II})
(lower frame). In both cases we have $\left\vert a\right\vert =1$ [see Eq. (%
\ref{|a|-RZ})], which leaves the excited state unpopulated in the end. For
several special values of the detuning $\Delta _{0}$, as predicted in Table %
\ref{Table-a}, we have $a=-1$. For these values, an equal superposition of
all degenerate states including the initially populated state $\left\vert
\psi _{1}\right\rangle $ is created in the upper frame, and an equal
superposition of all degenerate states except $\left\vert \psi
_{1}\right\rangle $ is created in the lower frame.

\begin{figure}[tb]
\includegraphics[width=75mm]{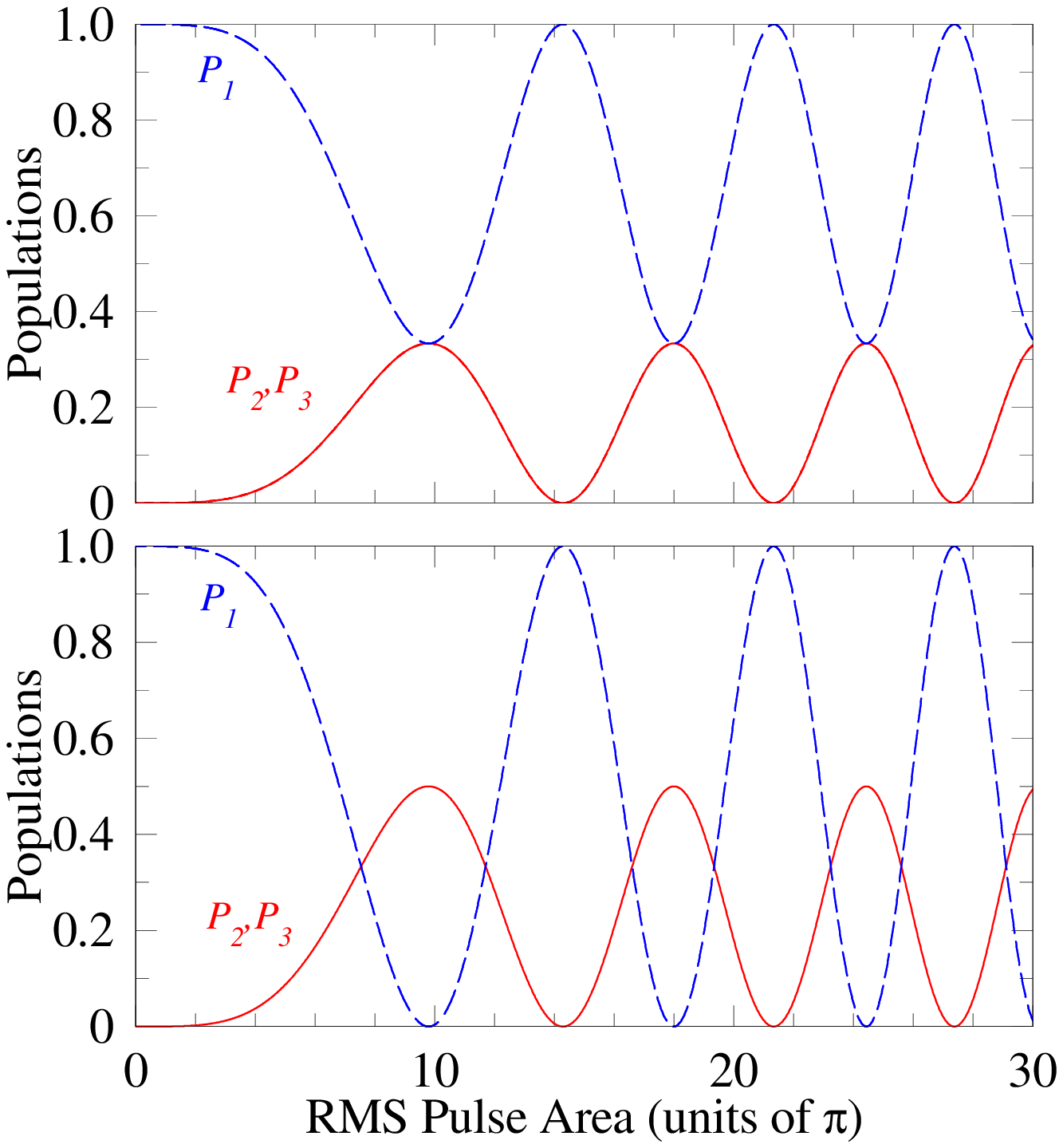}
\caption{(Color online) Final populations versus the rms pulse area $\protect%
\chi $ for $N=3$ degenerate lower states and detuning $\Delta T=50.534$. The
coupling strengths $\protect\chi _{n}$ are given by Eqs. (\protect\ref{Case
I}) in the upper frame and Eqs. (\protect\ref{Case II}) in the lower frame.
The system is initially in state $\left\vert \protect\psi _{1}\right\rangle $%
.}
\label{Fig-area}
\end{figure}

Figure \ref{Fig-area} shows the final populations versus the rms pulse area $%
A=\pi \chi T$ for $N=3$ degenerate lower states for couplings chosen to
satisfy Eqs. (\ref{Case I}) (upper frame) and (\ref{Case II}) (lower frame).
As follows from Table \ref{Table-a}, an equal superposition of all
degenerate states is created for rms pulse area $A=18\pi $; this is indeed
seen in the figure in the upper frame. For the same value of $A$ in the
lower frame an equal superposition is created of all degenerate states
except the initially populated state $\left\vert \psi _{1}\right\rangle $.
In both frames, there are other values of $A$ for which the same
superpositions are apparently created; a closer examination (not shown)
reveals that for these other values of the rms pulse area the created
superposition has almost, but not exactly, equal components.

\begin{figure}[tb]
\includegraphics[width=75mm]{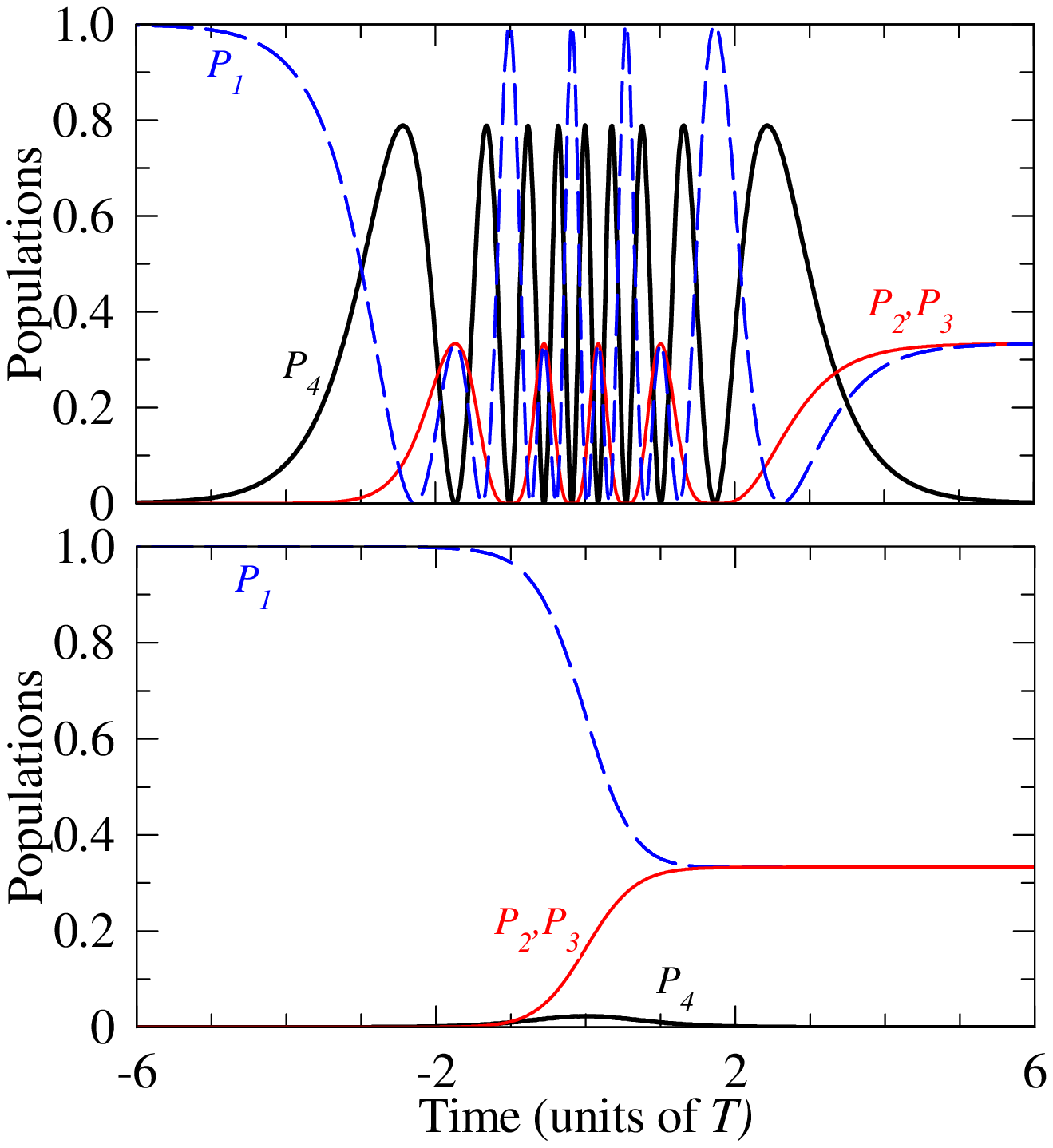}
\caption{(Color online) Populations versus time for $N=3$ lower states and
rms Rabi frequency $\protect\chi T=18$. The coupling strengths $\protect\chi %
_{n}$ are given by Eqs. (\protect\ref{Case I}). The detuning is $\Delta =0$
in the upper frame and $\Delta T=50.534$ in the lower frame. The system is
initially in state $\left\vert \protect\psi _{1}\right\rangle $.}
\label{Fig-time}
\end{figure}

Figure \ref{Fig-time} displays the time evolution of the populations for $%
N=3 $ degenerate lower states and rms pulse area of $18\pi $, and two
detunings: $\Delta =0$ in the upper frame and $\Delta T=50.534$ in the lower
frame. For these pairs of areas and detunings, Figs. \ref{Fig-detuning} and %
\ref{Fig-area} have already demonstrated that an equal superposition of all
degenerate states is created. Figure \ref{Fig-time} shows that the evolution
towards such a superposition can be dramatically different on and off
resonance. Indeed, for $\Delta =0$ (upper frame) the nondegenerate upper
state receives considerable transient population, which would lead to
significant losses if this state can decay on the time scale of the pulsed
interaction. In strong contrast, off resonance this undesired population is
greatly reduced (lower frame), and still the desired equal superposition of
the degenerate states emerges in the end. We have verified numerically that
for larger detunings this transient population continues to decrease, e.g.
for $\Delta T=142.198$ and $\Omega T=30$ it is less than 1\%.

To conclude this section we point out that one can create any desired
superposition, with arbitrary unequal populations, in very much the same
manner, on or off resonance, by appropriately chosing the individual
couplings, while still maintaning particular values of the overall rms pulse
area. Tuning on resonance gives the advantage of smaller pulse area
required, whereas tuning off resonance (with larger pulse area) provides the
advantage of greatly reducing the transient population of the possibly lossy
common upper state.

\section{Multistate Landau-Zener model\label{Sec-LZ}}

As seen in Table \ref{Table-a-exact} the propagator parameter $a$ for the
LZ\ model, $a=\exp \left( -\pi \chi ^{2}/4C\right) $, cannot be equal to 0
or 1 or $-1$, but may approach 0 or 1 arbitrarily closely. However, it is
always positive and cannot approach the value of $-1$; hence the LZ model is unsuitable for unitary
operations within the degenerate manifold, in contrast to the resonance and
RZ models discussed above. Still, the present multistate LZ\ solution
represents an interesting and important addition to the available LZ\
solutions (see \cite{CI transitions} and references therein).

\subsubsection{The Demkov-Osherov model}

The present multistate LZ model complements the Demkov-Osherov (DO) model
\cite{DO}, wherein a slanted energy crosses $N$ parallel \emph{nondegenerate}
energies. In the DO model, the exact probabilities $P_{n\rightarrow m}$ have
the same form --- products of LZ probabilities for transition or
no-transition applied at the relevant crossings --- as what would be
obtained by naive multiplication of LZ probabilities while moving across the
grid of crossings from $\left\vert \psi _{n}\right\rangle $ to $\left\vert
\psi _{m}\right\rangle $, without accounting for phases and interferences.
For example, if the states $\left\vert \psi _{n}\right\rangle $ ($%
n=1,2,\ldots ,N$) are labeled such that their energies increase with the
index $n$, and if the slope of the slanted energy of state $\left\vert \psi
_{N+1}\right\rangle $ is positive, the transition probabilities in the DO
model are
\begin{subequations}
\label{DO solution}
\begin{eqnarray}
P_{n\rightarrow m} &=&p_{n}q_{n+1}q_{n+2}\cdots q_{m-1}p_{m}\quad (n<m), \\
P_{n\rightarrow m} &=&0\quad (n>m), \\
P_{n\rightarrow n} &=&q_{n}, \\
P_{n\rightarrow N+1} &=&p_{n}q_{n+1}q_{n+2}\cdots q_{N}, \\
P_{N+1\rightarrow n} &=&q_{1}q_{2}\cdots q_{n-1}p_{n}, \\
P_{N+1\rightarrow N+1} &=&q_{1}q_{2}\cdots q_{N},
\end{eqnarray}%
\end{subequations}
where $q_{n}=\exp \left( -\pi \chi _{n}^{2}/2C\right) $ is the no-transition probability
 and $p_{n}=1-q_{n}$ is the transition probability between
states $\left\vert \psi _{N+1}\right\rangle $ and $\left\vert \psi
_{n}\right\rangle $ at the crossing of their energies.

\subsubsection{The degenerate case}

The present multistate LZ\ solution provides the transition probabilities
for the special case when all parallel energies are degenerate, which cannot
be obtained from the DO model.

\paragraph{The propagator}

The elements of the transition matrix for our $\left( N+1\right) $-state
degenerate LZ problem are readily found from Eq. (\ref{U}) to be
\begin{subequations}
\label{U-LZ}
\begin{gather}
U_{m,n}=-\frac{\chi _{n}\chi _{m}}{\chi ^{2}}\left( 1-e^{-\Lambda }\right)
\quad \left( m,n=1,\ldots ,N;\text{ }m\neq n\right) , \\
U_{n,n}=1-\frac{\chi _{n}^{2}}{\chi ^{2}}\left( 1-e^{-\Lambda }\right) \quad
\left( n=1,\ldots ,N\right) , \\
U_{n,N+1}=\frac{\chi _{n}}{\chi }b\quad \left( n=1,\ldots ,N\right) , \\
U_{N+1,n}=-\frac{\chi _{n}}{\chi }b^{\ast }\quad \left( n=1,\ldots ,N\right)
, \\
U_{N+1,N+1}=e^{-\Lambda },
\end{gather}%
with $\Lambda =\pi \chi ^{2}/4C$ and $\left\vert b\right\vert
^{2}=1-e^{-2\Lambda }$.

\paragraph{System initially in the nondegenerate state}

When the system begins initially in the nondegenerate state $\left\vert \psi
_{N+1}\right\rangle $, with the tilted energy, the system ends in a coherent
superposition of all states with populations
\end{subequations}
\begin{subequations}
\label{P-LZ-upper}
\begin{eqnarray}
P_{n} &=&\frac{\chi _{n}^{2}}{\chi ^{2}}\left( 1-e^{-2\Lambda }\right) \quad
\left( n=1,\ldots ,N\right) , \\
P_{N+1} &=&e^{-2\Lambda }.
\end{eqnarray}%
In the adiabatic limit $\Lambda \gg 1$ the population is distributed among
the degenerate states according to their couplings, whereas the initially
populated state $\left\vert \psi _{N+1}\right\rangle $ is almost depleted, $%
P_{N+1}\approx 0$. For equal couplings, all degenerate-state populations
will be equal, $P_{n}\approx 1/N$. In the opposite, diabatic limit $\Lambda
\ll 1$ the population remains in state $\left\vert \psi _{N+1}\right\rangle $
with almost no population in the degenerate states.

\paragraph{System initially in a degenerate state}

When the system is initially in an arbitrary degenerate state $%
\left\vert \psi _{i}\right\rangle $, at the end of the evolution the
populations are
\end{subequations}
\begin{subequations}
\label{P-LZ-lower}
\begin{eqnarray}
P_{i} &=&\left[ 1-\frac{\chi _{i}^{2}}{\chi ^{2}}\left( 1-e^{-\Lambda
}\right) \right] ^{2}, \\
P_{n} &=&\frac{\chi _{n}^{2}\chi _{i}^{2}}{\chi ^{4}}\left( 1-e^{-\Lambda
}\right) ^{2}\quad \left( n=1,\ldots ,N;n\neq i\right) , \\
P_{N+1} &=&\frac{\chi _{i}^{2}}{\chi ^{2}}\left( 1-e^{-2\Lambda }\right) .
\end{eqnarray}%
In the adiabatic limit $\Lambda \gg 1$ and for equal couplings, the
populations will be
\end{subequations}
\begin{subequations}
\label{LZ adiabatic}
\begin{eqnarray}
P_{i} &\approx &\left( 1-\frac{1}{N}\right) ^{2}, \\
P_{n} &\approx &\frac{1}{N^{2}}\quad \left( n=1,\ldots ,N;n\neq i\right) , \\
P_{N+1} &\approx &\frac{1}{N}.
\end{eqnarray}
\end{subequations}

Obviously, Eqs. (\ref{P-LZ-upper}) and (\ref{P-LZ-lower}) cannot be reduced
to the DO solution (\ref{DO solution}), which implies that the
non-degeneracy assumption in the DO model is essential.

\begin{figure}[t]
\includegraphics[width=75mm]{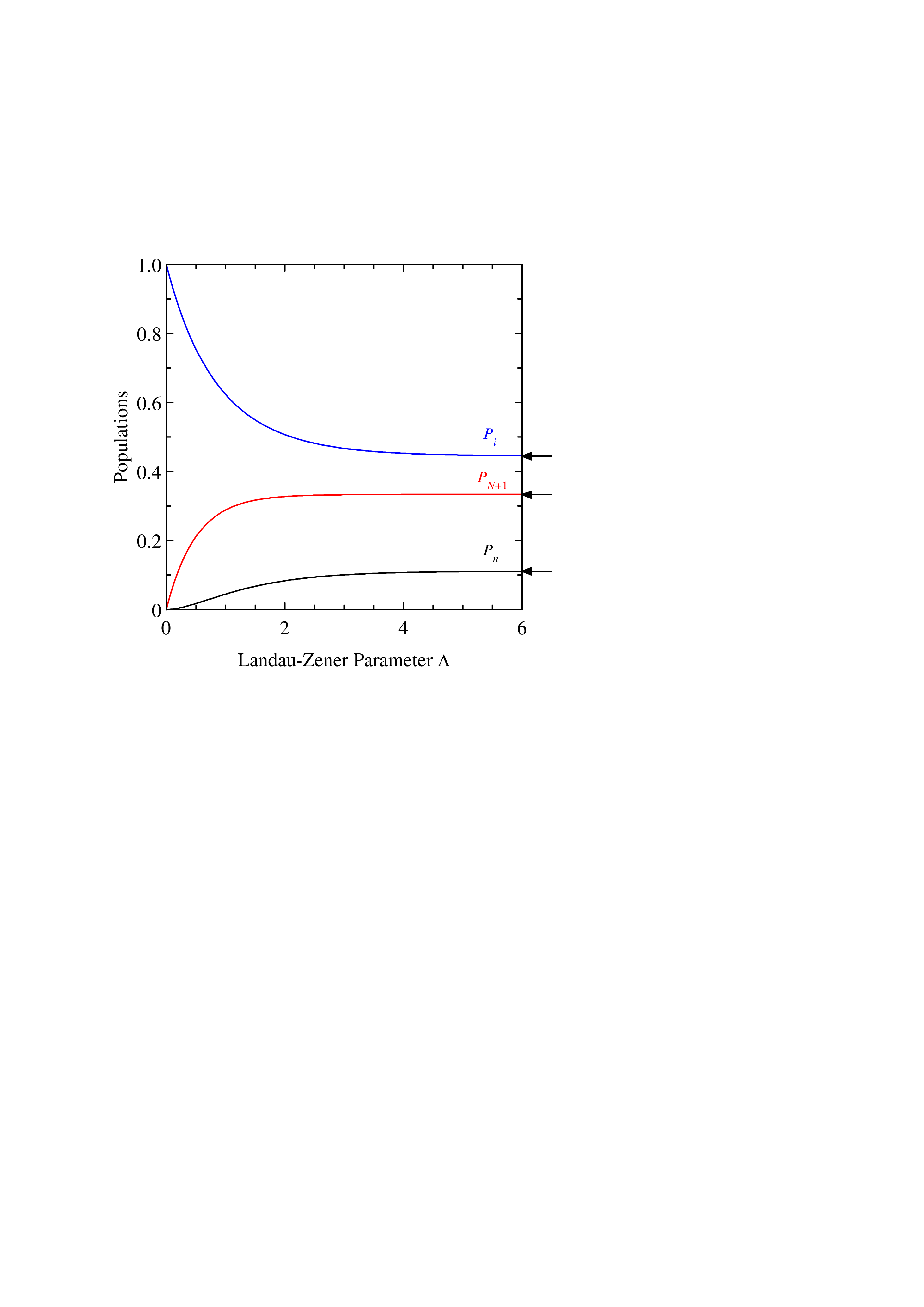}
\caption{(Color online) Populations for the degenerate LZ\ model vs the LZ
parameter $\Lambda =\protect\pi \protect\chi ^{2}/4C$ for $N=3$ degenerate
states and equal couplings. The system is supposed to start in one of the
degenerate states $\left\vert \protect\psi _{i}\right\rangle $. The arrows
on the right point the adiabatic values (\protect\ref{LZ adiabatic}).}
\label{Fig-LZ}
\end{figure}

Figure \ref{Fig-LZ} shows the transition probability for the multistate LZ
model plotted against the LZ parameter $\Lambda =\pi \chi ^{2}/4C$. As $%
\Lambda $ increases the populations approach their steady adiabatic values (%
\ref{LZ adiabatic}). Different coherent superpositions can be created by
choosing appropriate values for the couplings $\chi _{n}$.

\section{Conclusions\label{Sec-conclusions}}

In this paper we have described a procedure for deriving analytical
solutions for a multistate system composed of $N$ degenerate lower states
coupled via a nondegenerate upper state with pulsed interactions of the same
temporal dependence but possibly with different peak amplitudes. The
multistate resonance and Rosen-Zener solutions have been discussed in some
detail because they allow one to find special values of parameters, termed
generalised $\pi $ pulses, for which various types of population transfer
can occur, for example, creation of maximally coherent superpositions. The
RZ solution is particularly useful because it allows to prescribe
appropriately detuned pulsed fields for which the dynamics can be
essentially contained within the degenerate-state space, without populating
the upper state even transiently, thus avoiding possible losses from this
state via spontaneous emission, ionization, etc.

We have analyzed in some detail also the multistate Landau-Zener
 model, which complements the Demkov-Osherov model in the case of degenerate energies.

The presented analytical solutions and general properties
have a significant potential for manipulation of multistate quantum bits
 in quantum information processing, for example, in designing
 arbitrary unitary gates.

\acknowledgments

This work has been supported by the European Union's Transfer of Knowledge
project CAMEL (Grant No. MTKD-CT-2004-014427) and the Alexander von Humboldt
Foundation. ESK acknowledges support from the EU Marie Curie Training Site
project No. HPMT-CT-2001-00294.


\begin{thebibliography}{99}
\bibitem{Shore} B.W. Shore, \emph{The Theory of Coherent Atomic Excitation}
(Wiley, New York, 1990).

\bibitem{Rabi} I.I. Rabi, Phys. Rev. \textbf{51}, 652 (1937).

\bibitem{LZ} L.D. Landau, Physik Z. Sowjetunion \textbf{2}, 46 (1932); C.
Zener, Proc. R. Soc. Lond. Ser. A \textbf{137}, 696 (1932).

\bibitem{RZ} N. Rosen and C. Zener, Phys. Rev. \textbf{40}, 502 (1932).

\bibitem{AE} L. Allen and J. H. Eberly, \emph{Optical Resonance and
Two-Level Atoms} (Dover, New York, 1987); F.T. Hioe, Phys. Rev. A \textbf{30}%
, 2100 (1984).

\bibitem{BB} A. Bambini and P.R. Berman, Phys. Rev. A \textbf{23}, 2496
(1981).

\bibitem{DK} Yu.N. Demkov and M. Kunike, Vestn. Leningr. Univ. Fiz. Khim.
\textbf{16}, 39 (1969); see also F.T. Hioe and C.E. Carroll, Phys. Rev. A
\textbf{32}, 1541 (1985); J. Zakrzewski, Phys. Rev. A \textbf{32}, 3748
(1985); K.-A. Suominen and B.M. Garraway, Phys. Rev. A \textbf{45}, 374
(1992).

\bibitem{CH} C.E. Carroll and F.T. Hioe, J. Phys. A: Math. Gen. \textbf{19},
3579 (1986).

\bibitem{Demkov} Yu.N. Demkov, Sov.Phys.-JETP \textbf{18}, 138 (1964); N.V.
Vitanov, J. Phys. B \textbf{26}, L53 (1993), erratum \emph{ibid.} \textbf{26}%
, 2085 (1993).

\bibitem{Nikitin} E.E. Nikitin, Opt. Spectrosc. \textbf{13}, 431 (1962);
Discuss. Faraday Soc. \textbf{33}, 14 (1962); Adv. Quantum Chem. \textbf{5},
135 (1970); N.V. Vitanov, J. Phys. B \textbf{27}, 1791 (1994).

\bibitem{Morris-Shore} J.R. Morris and B.W. Shore, Phys. Rev. A \textbf{27},
906 (1983).

\bibitem{QI} C.P. Williams and S.H. Clearwater, \emph{Explorations in
Quantum Computing}, (Springer-Verlag, Berlin, 1997); A. Steane, Rep. Prog.
Phys. \textbf{61}, 117 (1998); M.A. Nielsen and I.L. Chuang, \emph{Quantum
Computation and Quantum Information} (Cambridge University Press, Cambridge,
2000).

\bibitem{tripod} R.G. Unanyan, M. Fleischhauer, B.W. Shore, and K. Bergmann,
Opt. Commun. \textbf{155}, 144 (1998); H. Theuer, R.G. Unanyan, C.
Habscheid, K. Klein and K. Bergmann, Optics Express \textbf{4}, 77 (1999).

\bibitem{Kis} Z. Kis and S. Stenholm, Phys. Rev. A \textbf{64}, 63406 (2001).

\bibitem{STIRAP} K. Bergmann, H. Theuer, and B.W. Shore, Rev. Mod. Phys.
\textbf{70}, 1003 (1998); N.V. Vitanov, T. Halfmann, B.W. Shore, and K.
Bergmann, Ann. Rev. Phys. Chem. \textbf{52}, 763 (2001); N.V. Vitanov, M.
Fleischhauer, B.W. Shore and K. Bergmann, Adv. At. Mol. Opt. Phys. \textbf{46%
}, 55 (2001).

\bibitem{Stenholm} S. Stenholm, in \emph{Frontiers of Laser Spectroscopy},
Les Houches Summer School Session XXVII, edited by R. Bailian, S. Haroche
and S. Liberman (Amsterdam, North Holland, 1975), p. 399;
 R. Lefebvre and J. Savolainen, J. Chem. Phys. \textbf{60}%
, 2509 (1974); M. Bixon and J. Jortner, J. Chem. Phys. \textbf{48}, 715
(1968).
\bibitem{Vitanov98} N.V. Vitanov, J. Phys. B \textbf{33}, 2333 (2000).

\bibitem{AS} M. Abramowitz and I.A. Stegun, \textit{Handbook of Mathematical
Functions} (Dover, New York, 1964).

\bibitem{CI transitions} A.A. Rangelov, J. Piilo, and N.V. Vitanov, Phys.
Rev. A \textbf{72}, 053404 (2005).

\bibitem{DO} Y.N. Demkov and V.I. Osherov, Zh. Eksp. Teor. Fiz. \textbf{53},
1589 (1967) [Sov. Phys. JETP \textbf{26}, 916 (1968)]; Y.N. Demkov and V.N.
Ostrovsky, J. Phys. B \textbf{28}, 403 (1995).

\end{thebibliography}
\end{document}